# Beyond the Maxwell Limit: Thermal Conduction in Nanofluids with Percolating Fluid Structures

Jacob Eapen*, Ju Li+ and Sidney Yip++

*Theoretical Division, Los Alamos National Laboratory, Los Alamos, NM 875445 (eapen@lanl.gov).
+Department of Materials Science and Engineering, Ohio State University, Columbus, OH 43210.
++Department of Nuclear Science and Engineering, Massachusetts Institute of Technology, Cambridge, MA 02139.

## Abstract

*In a well-dispersed nanofluid with strong cluster-fluid attraction, thermal conduction paths can arise through percolating amorphous-like interfacial structures. This results in a thermal conductivity enhancement beyond the Maxwell limit of* $3\phi$*, with* $\phi$ *being the nanoparticle volume fraction. Our findings from non-equilibrium molecular dynamics simulations, which are amenable to experimental verification, can provide a theoretical basis for the development of future nanofluids.*

Mean-field theories are largely successful for characterizing the thermal conduction behavior in nanofluids [1-3]. In particular, the thermal conductivity of most nanofluids is well-characterized by the dilute limit of $3\phi$ from the classical Maxwell's theory for well-dispersed, spherical nanoparticles, with $\phi$ denoting the nanoparticle volume fraction [1]. Several recent experiments, however, show conflicting behavior such as increasing thermal conductivity with decreasing nanoparticle size [4], saturation at higher volume fractions [5], an apparent lack of correlation to the intrinsic thermal conductivity of the nanoparticles [5, 6], and a relatively large thermal conductivity enhancement at low volume fractions (18-25% with $\phi \leq 1\%$ )[5, 7].

From a theoretical perspective, a key challenge is to understand the possible mechanisms that can increase the thermal conductivity beyond the $3\phi$ Maxwell limit. The postulated Brownian-motion or micro-convection mechanisms [8] have been shown to be untenable through experimental [9] and theoretical studies [10]. Two recent studies, a modification of the mean-field theory [11] and our equilibrium molecular dynamics simulation of small clusters [12], open up interesting theoretical possibilities. In the first, the thermal conductivity is shown to increase by percolation through a chain-like agglomeration of the nanoparticles. In the second, we find an enhancement in the thermal conductivity originating from the correlation in the potential energy flux in the presence of strong cluster-fluid attraction.

In this Letter, we demonstrate using non-equilibrium molecular dynamics simulations (NEMD) that the thermal conductivity of a well-dispersed nanofluid is enhanced beyond the $3\phi$ Maxwell limit through a percolating amorphous-like *fluid* structure at the cluster interface. These interconnected paths emerge only when the cluster-fluid interaction is strong; for weak cross-interaction the enhancement is only slightly higher than the $3\phi$ limit. The attendant changes in interfacial structure are accessible by experimental techniques such as neutron scattering. Our findings suggest a theoretical basis for a systematic development of future nanofluids.

We consider a nanofluid system where solid clusters are dispersed uniformly in a Lennard-Jones (LJ) fluid. The cluster atoms, in addition to experiencing a LJ potential, are held together by a finitely extensible nonlinear elastic (FENE) potential given by $-A\varepsilon \ln[1-(r/B\sigma)^2]$ where $\varepsilon$ and $\sigma$ are the reference values for energy and length, and $A$ and $B$ are two constants which take the values of 5.625 and 4.95 respectively in the simulations. The fluid-fluid and solid-fluid (SF) interactions are modeled by LJ interactions with the parameters ($\varepsilon$, $\sigma$) and ($\varepsilon_{SF}$, $\sigma$) respectively. The coupling between the solid and fluid atoms is measured through the attractive potential well-depth $\varepsilon_{SF}$ relative to that between two fluid atoms, $\varepsilon$. Reduced units based on $m$, $\varepsilon$ and $\sigma$ are used throughout in this study. Our simulations are carried out at a constant temperature of 1.0 and a volume corresponding to zero pressure. At this state point, the radial distribution function (*rdf*) indicates that the base fluid has a structure corresponding to that of a liquid. The cluster configuration consists of ten solid clusters of 10 atoms each in a host fluid of 1948 atoms leading to a number fraction of 4.88%. The volume fraction is calculated as $\pi n\rho a^3/6v$, where $n$ is the solid atom number fraction, $\rho$ is the number density, and $a$ and $v$ are the nearest neighbor distance and the packing fraction in a FCC lattice respectively. The ratio $a^3/v$ assigns the void space between the solid atoms to the cluster volume.

NEMD simulation is conducted through the imposed-flux method [13] where a temperature profile develops with the imposition of a known heat flux. This method is compatible with periodic boundary conditions, conserves both energy and momentum, and experiences only limited perturbation effects. The simulation box is divided into slabs perpendicular to a chosen direction, say, $z$ with the edge slabs denoted as 'cold' and the center slab as 'hot'. Periodic velocity exchanges are made between atoms of these slabs such that the hottest atom in the cold slab is exchanged with the coldest atom in the hot slab. At steady state, a linear temperature profile develops which is symmetric about the hot slab. The heat flux is computed exactly with the known values of the velocities that are exchanged [13]. The equilibration is done for 100 000 iterations and temperature in each slab is averaged for another 150 000 iterations without the use of a thermostat or barostat. Further averaging over 10-12 initial conditions are needed to average out the statistical variations of the clusters and to generate an acceptable linearity in the temperature profile (measured by the multiple correlation coefficient, $R^2 \geq 0.999$). No significant size dependency has been observed in the simulation results. The thermal conductivity is evaluated from the Fourier's law $\kappa = -\langle \mathbf{J}_q \rangle_z / (dT/dz)$ , where $\mathbf{J}_q$ and $dT/dz$ are the heat flux and temperature gradient respectively.

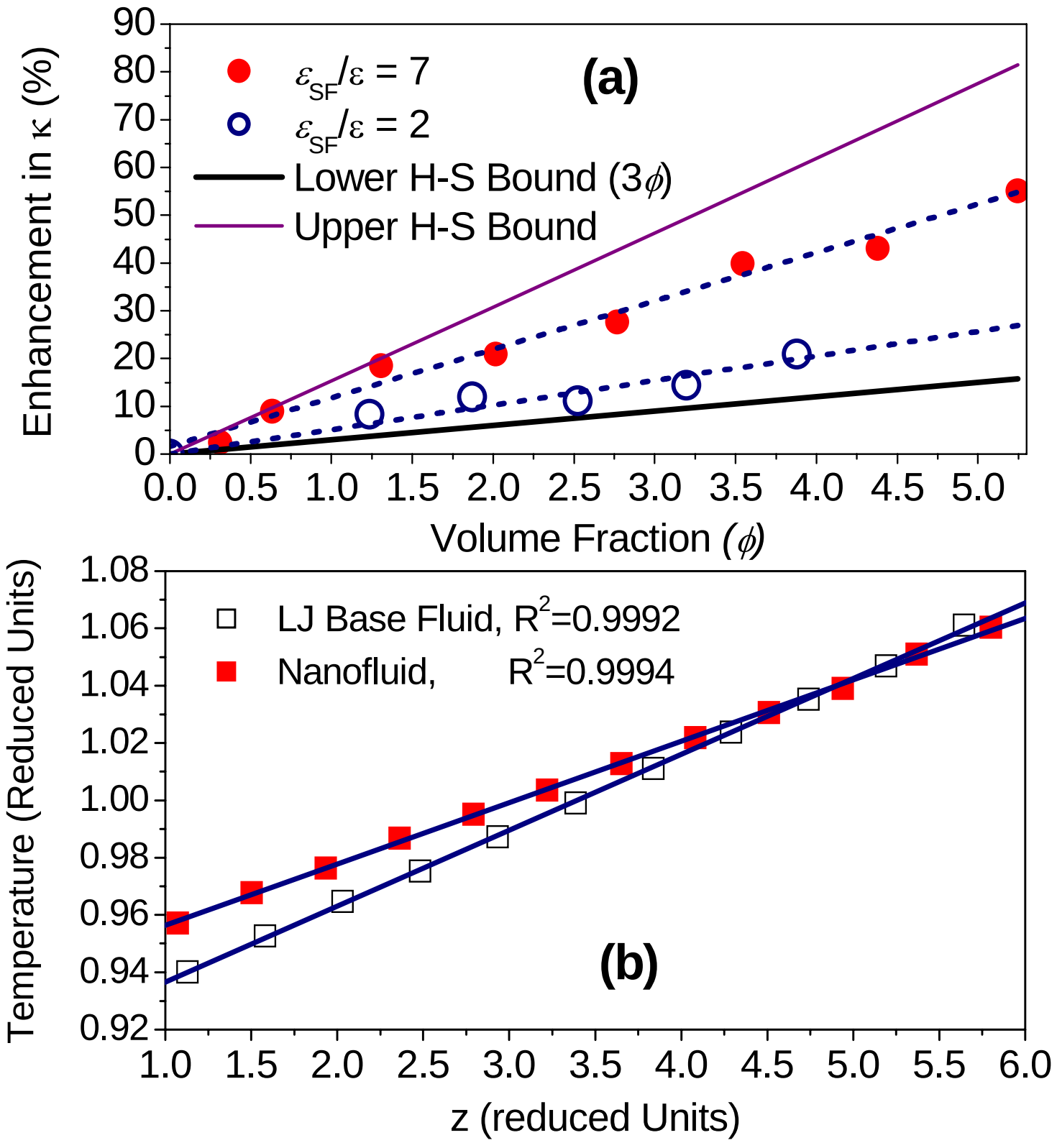

Fig. 1. (a) Enhancement in $\kappa$ for $\varepsilon_{SF}/\varepsilon$=7 and $\varepsilon_{SF}/\varepsilon$=2 at $T$=1 and $P$=0, and the Hashin-Shtrikman (H-S) bounds. (b) Temperature profiles with $\varepsilon_{SF}/\varepsilon$=7 for $\phi = 3.5\%$ . $R^2$ denote the multiple correlation coefficient.

In Figure 1 (a) we show that the $\kappa$ enhancements with the weak ($\varepsilon_{SF}/\varepsilon$=2) as well as the strong cluster-fluid attraction ($\varepsilon_{SF}/\varepsilon$=7) increase linearly with $\phi$ . The thermal conductivities are evaluated from well-converged temperature profiles as shown in Figure 1(b). For the strong cross-attraction, $\kappa$ increases with a much larger slope than that of the weak cross-attraction, resulting in an enhancement of 55% at volume fraction of 5.2%.

The simulation data, however, are bounded by the formal limits of Hashin-Shtrikman (H-S) mean-field theory [14] wherein the lower limit is identically equivalent to the Maxwell expression for $\kappa_p > \kappa_f$ where $\kappa_p$ and $\kappa_f$ are the nano-cluster and fluid thermal conductivities. The H-S expression is given by:

$$\kappa_f \left(1+\frac{3\phi[\kappa]}{3\kappa_f+(1-\phi)[\kappa]}\right) \le \kappa \le \left(1-\frac{3(1-\phi)[\kappa]}{3\kappa_p-\phi[\kappa]}\right)\kappa_p \qquad (1)$$

where $[\kappa]$ is defined as $\kappa_p - \kappa_f$. For the weaker cross-attraction, the enhancements are only slightly larger than the lower H-S bound (Maxwell $3\phi$ limit) while for the stronger cross-attraction, they are closer to the upper H-S bound. The H-S theory does not give the actual mechanism of thermal conductivity enhancement but sets the most restrictive bounds on the basis of knowing only the volume fraction. Physically, the upper H-S bound corresponds to a nano-cluster matrix with spherical inclusions of fluid regions, and the lower H-S bound is for the reverse situation. The upper bound, therefore, is skewed towards a parallel conduction mode between the nano-clusters and fluid atoms, whereas the lower bound favors a series mode. So it is evident that the strong cross-attraction introduces conduction paths that are parallel and since the clusters are well dispersed, they are possible only through the mediation of the fluid atoms. We will now explain this behavior by analyzing the interfacial fluid structure.

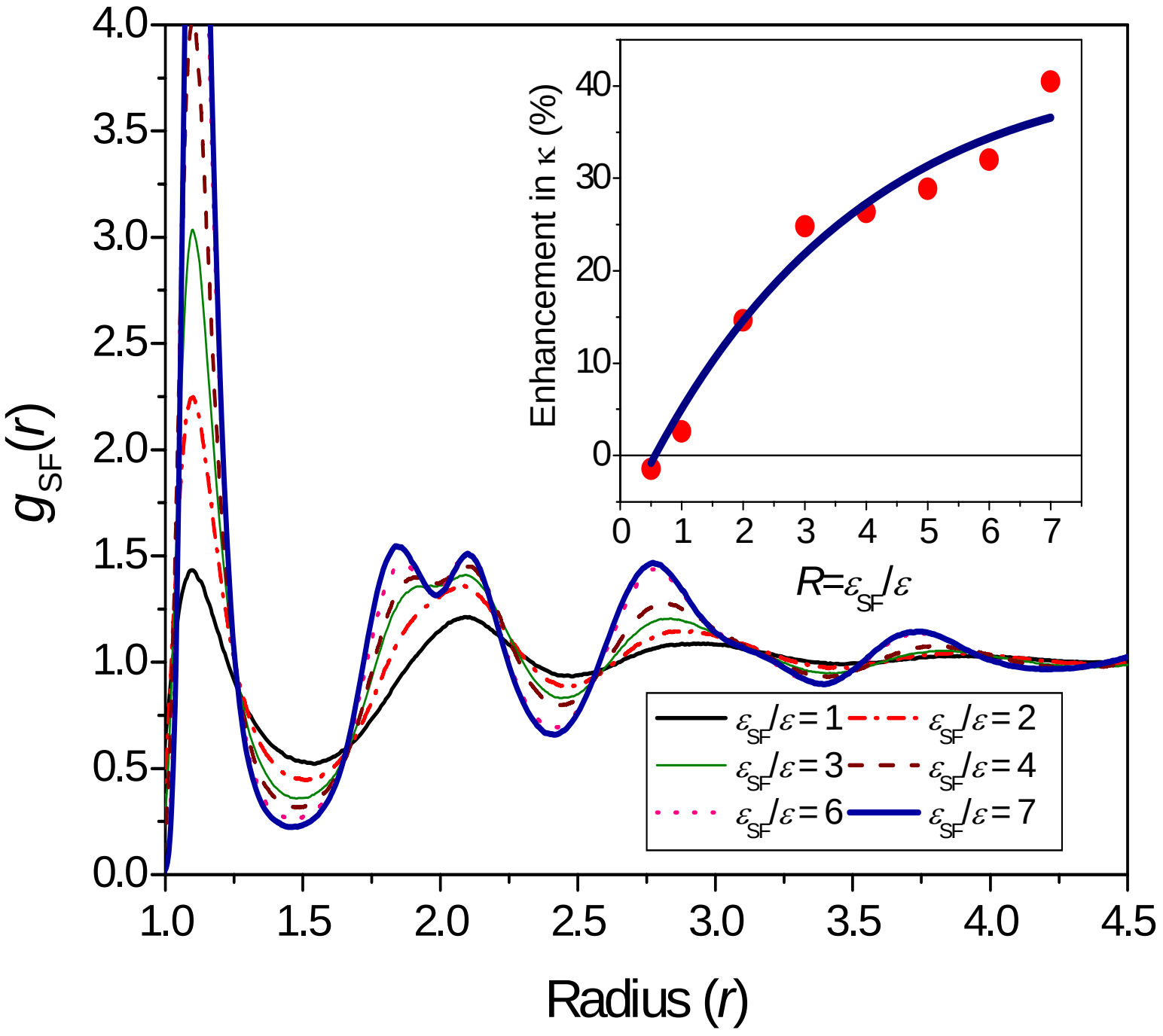

Fig. 2. The cross radial distribution function $g_{SF}$(r) for several values of $\varepsilon_{SF}/\varepsilon$ with $n$=100/2048. Inset shows the increase in the thermal conductivity with $\varepsilon_{SF}/\varepsilon$.

In Figure 2, we show the cross radial distribution function, $g_{SF}$(r) for several values of $\varepsilon_{SF}/\varepsilon$ at a volume fraction of 3.5%. For $\varepsilon_{SF}/\varepsilon$=2, the cross radial distribution function, which is the probability of finding a fluid atom at a certain radial location given that a solid atom is at the origin, has the characteristics resembling that of a fluid. As $\varepsilon_{SF}/\varepsilon$ further increases, the fluid atoms in the vicinity of the clusters get more strongly attracted towards the clusters and move inwards, as can be seen from the sharpening of the first peak and the emergence of the third peak. Remarkably, the second peak flattens and splits into two, which is a characteristic signature of an amorphous-like transition. The fluid-fluid radial distribution function, on the other hand, is only marginally affected by the higher values of $\varepsilon_{SF}/\varepsilon$ (not shown). It is thus clear that the fluid atoms pack themselves in a random close pack arrangement only near the vicinity of the solid clusters. Inset (a) shows the enhancement in the thermal conductivity which increases with $\varepsilon_{SF}/\varepsilon$. Interestingly, a change of slope can be detected beyond $\varepsilon_{SF}/\varepsilon$=3 which is commensurate with a more modest change in the structure of the interfacial fluid atoms.

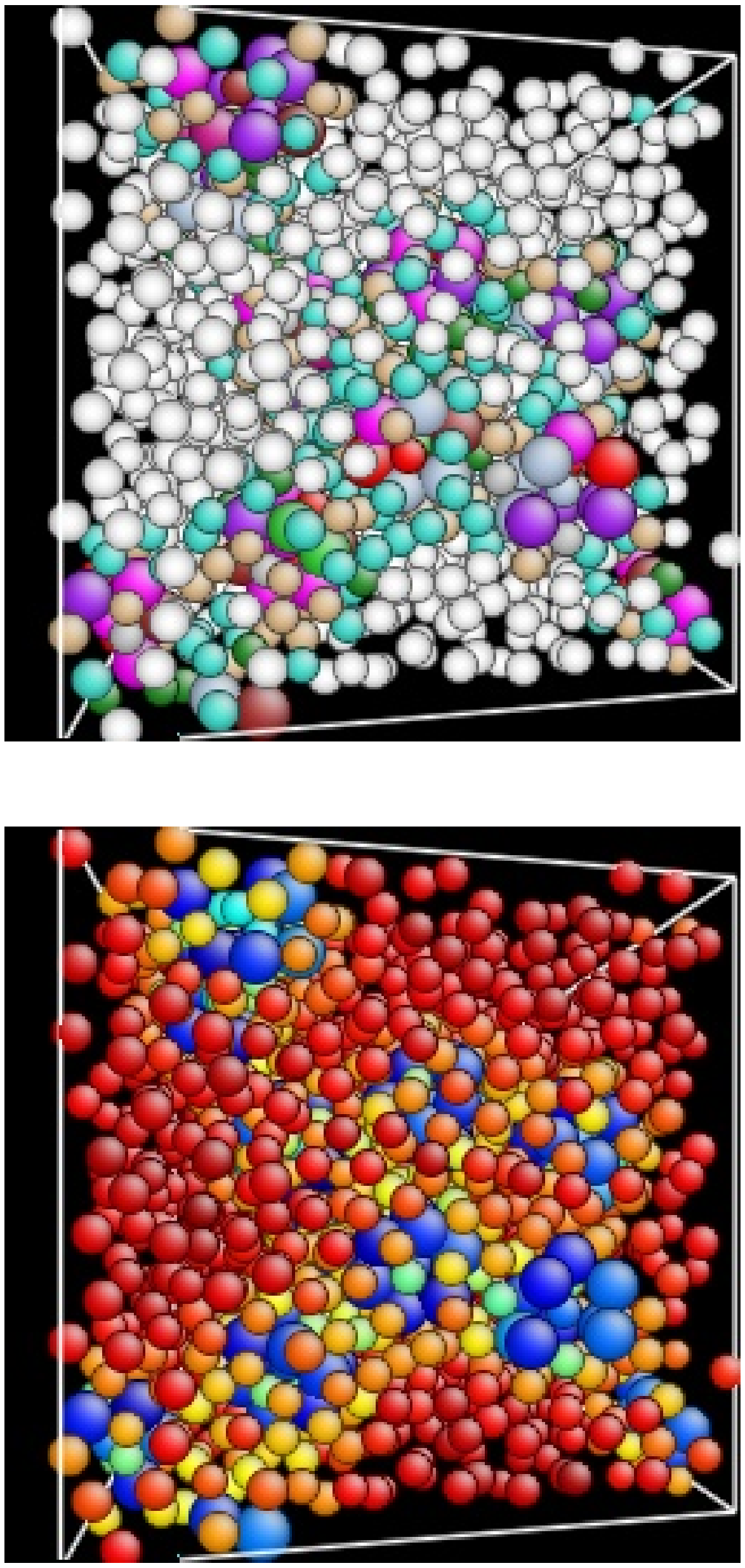

Fig. 3 (top) Coordination number (CN) and potential energy (bottom) maps with $\varepsilon_{SF}/\varepsilon$=7 and $\phi = 3.5\%$. Magenta/deep-blue and white/red in the CN/potential energy maps denote the (ten) clusters and fluid atoms respectively. The intermediate colors represent the percolating conduction paths.

Thus, a relatively strong cluster-fluid attraction ($\varepsilon_{SF}/\varepsilon \geq 2$) introduces an amorphous-like fluid structure around the clusters. The mean separation between the clusters is $O(d)$ as given by the dilute limit, $d_s = [(\pi/6\phi)^{1/3} - 1]d$, where $d$ is the cluster diameter, and $d_s$ the separation distance. This shows that the interface around the clusters forms a percolating network of amorphous-like structure. The co-ordination number map in Figure 3 (top), colored by the number of nearest neighbors, shows this percolating interfacial fluid structure. The higher density of fluid atoms along this structure allows a network of excess potential energies as shown in Figure 3 (bottom) which are instrumental in additional thermal conduction paths either through a potential energy exchange mechanism or a phonon-like collision mechanism [12]. The networked paths now favor a parallel mode of conduction and the observed enhancements in the thermal conductivity are closer to the upper H-S bound as observed in Figure 1.

In conclusion, we find that a strong cross-attraction between the clusters and fluid atoms in a nanofluid induces a percolating network of thermal conduction paths mediated by the interfacial fluid atoms. As a result, the thermal conductivity can be enhanced beyond the commonly observed $3\phi$ Maxwell limit.

**Acknowledgements**

The work of J.E. and S.Y. is supported by the National Science Foundation under Grant No. 0205411. The work of J.L. is supported by DOE DE-FG02-06ER46330.